\documentclass[useAMS,usenatbib,fleqn]{mn2e}
\usepackage{amsmath,amssymb}
\usepackage{graphics}
\usepackage{tabularx}
\usepackage{color}

\def\eqref#1{equation~(\ref{#1})}

\newcommand{\prog}[1]{\texttt{\lowercase{#1}}}

\newsavebox{\boxcmdbox}
\newenvironment{lcmd}{\setlength{\fboxsep}{0pt}\begin{lrbox}{\boxcmdbox}\hspace*{6mm}\begin{minipage}{160mm}\vspace*{2mm}\tt}{\vspace*{2mm}\end{minipage}\hspace*{5mm}\end{lrbox}\begin{center}\framebox{\colorbox[gray]{0.9}{\usebox{\boxcmdbox}}}\end{center}}

\newenvironment{ccmd}{\setlength{\fboxsep}{0pt}\begin{lrbox}{\boxcmdbox}\hspace*{4mm}\begin{minipage}{60mm}\vspace*{2mm}\tt}{\vspace*{2mm}\end{minipage}\hspace*{4mm}\end{lrbox}\framebox{\colorbox[gray]{0.9}{\usebox{\boxcmdbox}}}}
\def\ptab{\hspace*{10ex}}

\title[The FITSH package]{FITSH -- a software package for image processing}
\author[P\'al, A.]{Andr\'as P\'al%
\thanks{E-mail address: apal@szofi.net} \\
Konkoly Observatory of the Hungarian Academy of Sciences, 
        Konkoly Thege Mikl\'os \'ut 15-17,
        Budapest, H-1121, Hungary\\
Department of Astronomy, Lor\'and E\"otv\"os University, 
	P\'azm\'any P\'eter s\'et\'any 1/A, 
	Budapest H-1117, Hungary}

\def\fitsh{\texttt{FITSH}}

\begin{document}

\date{Accepted ..., Received ... ; in original form ...}

\pagerange{\pageref{firstpage}--\pageref{lastpage}} \pubyear{2010}

\maketitle

\label{firstpage}

\begin{abstract}
In this paper we describe the main features of the software package named
\fitsh{}, intended to provide a standalone environment for 
analysis of data acquired by imaging astronomical detectors.
The package provides utilities both for the full pipeline
of subsequent related data processing steps (incl. image calibration,
astrometry, source identification, photometry, differential analysis,
low-level arithmetic operations, multiple image combinations, 
spatial transformations and interpolations, etc.) and for aiding the 
interpretation of the (mainly photometric and/or astrometric) results. 
The package also features a consistent implementation of photometry based 
on image subtraction, point spread function fitting and aperture photometry and provides
easy-to-use interfaces for comparisons and for picking the most suitable
method for a particular problem. This set of utilities found in the package
are built on the top of the commonly used UNIX/POSIX shells (hence 
the name of the package), therefore
both frequently used and well-documented tools for such environments
can be exploited and managing massive amount of data is rather convenient.
\end{abstract}



\section{Introduction}
\label{sec:introduction}

In general, homogeneous data acquisition and subsequent data processing
are essential for obtaining a proper measurement and characterize
various observations. It is especially true for observing weak
signals and analyzing related data series with a relatively low
signal-to-noise (S/N) ratio. These mentioned data processing steps
include 
	a) calibration and per-pixel arithmetic operations;
	b) source detection, source profile characterization and (optionally
		space-varying) PSF determination;
	c) catalogue and cross-matching, coordinate list processing and astrometry;
	d) image registration, convolution and combination of multiple images;
	e) instrumental photometry (normal, PSF fitting and based 
		on image subtraction);
	f) refined profile modelling, obtaining centroid and shape parameters;
	g) creation of model and artificial images;
	h) data modelling and regression analysis.
Although several software solutions exist for processing imaging 
astronomical data (see e.g. IRAF\footnote{http://iraf.noao.edu}, 
ISIS\footnote{\cite{alard1998,alard2000}}, 
SExtractor\footnote{\cite{bertin1996}}, utilities and wrappers around
these implemented in Python\footnote{http://www.stsci.edu/resources/software\_hardware/pyraf}
or IDL), we intended to develop a lightweight package that is both a 
versatile solution for astronomical image processing (focusing
mostly on the processing of optical imaging data)
and features the many of the most recent data analysis and interpretation
algorithms. 

This paper presents a software package named \fitsh{}, that provides a 
set of independent binary programs 
(called ``tasks'') that are designed to be executed from a UNIX 
command line shell or shell script. Each of these tasks performs a specific 
operation (see the various steps above),
while the details of a certain operation are specified via command line 
switches and/or arguments. Therefore this package does not need any higher 
level operating environment than a standard UNIX shell, however, 
processing the related data might require a little more knowledge of 
the used shell itself (the documentation and available examples 
use the \texttt{bash} shell\footnote{http://www.gnu.org/software/bash/}). 
Additionally, some of the processing steps might 
require minor or basic operations performed with other tools like 
\texttt{awk}\footnote{http://www.gnu.org/software/gawk/}
or text processing utilities (\texttt{sort}, \texttt{uniq}, 
\texttt{paste}\footnote{http://www.gnu.org/software/coreutils/}, \dots). 
Another advantage of a ``plain'' UNIX environment is the option for exploiting
other shell-level features, such as very easy implementation
of remote execution, job control, batched processing (including background
processing),
higher level ways of run tasks in 
parallel\footnote{see e.g. http://www.gnu.org/software/pexec/},
and integration for autonomous observing 
systems\footnote{Some of such applications of this package are planned 
to be described later on in further papers.}

In order to have a consistent implementation of the procedures
required by astronomical image processing, several well known algorithms
(that are ultimately used as standard procedures) have also been implemented
in the \fitsh{} package. The details of these are known from the literature.
These include image preprocessing and calibration \citep{chromey2010},
basic source extraction, aperture photometry and PSF modelling 
\citep{stetson1987,stetson1989} or differential image 
analysis \citep{alard1998}. The new improvements, including 
routines related to astrometry and catalogue matching, image
transformations, consistent photometry on subtracted (differential) images, 
various regression analysis methods, etc. have been discussed in more details 
in \cite{pal2009phd}. See this work and references therein for
more details. In this paper we also refer to some parts of that thesis
during the discussion of various tasks. 

The structure of this paper goes as follows. Sec.~\ref{sec:implementation}
describes briefly some aspects of the practical implementation.
In Sec.~\ref{sec:tasks}, each of the main tasks are described
in more detail, featuring small ``script-lets'' as a demonstration of 
the syntax of the various tasks. 
In Sec.~\ref{sec:discussion}, we summarize the results.
The software package has its own website located at 
\texttt{http://fitsh.szofi.net}. This site displays information about the 
program and the tasks (including documentation and detailed examples), 
it is the primary download source of the package itself, and additionally, 
serves a public forum for the program users in various topics.


\section{Implementation aspects}
\label{sec:implementation}

\begin{table*}
\caption{An overview of the standalone tasks included 
in the package, displaying their main purposes and the types of input and
output data.
\label{tab:progsummary}}
\begin{center}
\sloppy
\begin{tabularx}{174mm}{|l|X|X|X|}
\hline 
Task & Main purpose & Type of input & Type of output \\
\hline
\texttt{fiarith}
&
Evaluates arithmetic expressions on images as operands.
&
A set of FITS images.
& 
A single FITS image.
\\
\texttt{ficalib}
&
Performs various calibration steps on the input images.
&
A set of raw FITS images.
& 
A set of calibrated FITS images.
\\
\texttt{ficombine}
&
Combines (most frequently averages) a set of images.
&
A set of FITS images.
& 
A single FITS image.
\\
\texttt{ficonv}
&
Obtains an optimal convolution transformation between two images
or use an existing convolution transformation to convolve an image.
&
Two FITS images or a single image and a transformation.
&
A convolution transformation or a single image.
\\
\texttt{fiheader}
&
Manipulates, i.e. reads, sets, alters or removes some FITS header keywords
and/or their values.
&
A single FITS image (alternation) or more FITS images (if header contents
are just read).
&
A FITS image with altered header or a series of keywords/values from the
headers.
\\
\texttt{fiign}
&
Performs low-level manipulations on masks associated to FITS images.
&
A single FITS image (with some optional mask).
&
A single FITS image (with an altered mask).
\\
\texttt{fiinfo}
&
Gives some information about the FITS image in a human-readable form
or creates image stamps in a conventional format.
&
A single FITS image.
&
Basic information or PNM images.
\\
\texttt{fiphot}
&
Performs photometry on normal, convolved or subtracted images.
&
A single FITS image (with additional reference photometric information
if the image is a subtracted one).
&
Instrumental photometric data.
\\
\texttt{firandom}
&
Generates artificial object lists and/or artificial (astronomical) images.
&
List of sources to be drawn to the image or an arithmetic expression that 
describes how the list of sources is to be created.
&
List of sources and/or a single FITS image.
\\
\texttt{fistar}
&
Detects and characterizes point-like sources from astronomical images.
&
A single FITS image.
&
List of detected sources and an optional PSF image (in FITS format).
\\
\texttt{fitrans}
&
Performs generic geometric (spatial) transformations on the input image.
&
A single FITS image.
&
A single, transformed FITS image.
\\
\texttt{fi[un]zip}
&
Compresses and decompresses primary FITS images.
&
A single uncompressed or compressed FITS image file.
&
A single compressed or uncompressed FITS image file.
\\
\texttt{grcollect}
&
Performs data transposition on the input tabulated data or do some sort
of statistics on the input data.
&
A set of files containing tabulated data.
&
A set of files containing the transposed tabulated data or a single
file for the statistics, also in a tabulated form.
\\
\texttt{grmatch}
&
Matches lines read from two input files of tabulated data, using
various criteria (point matching, coordinate matching or identifier matching).
&
Two files containing tabulated data (that must be two point sets in the
case of point or coordinate matching).
&
One file containing the matched lines and in the case of point matching,
an additional file that describes the best fit geometric transformation between
the two point sets.
\\
\texttt{grselect}
&
Selects lines from tabulated data using various criteria.
&
A single file containing tabulated data.
&
The filtered rows from the input data.
\\
\texttt{grtrans}
&
Transforms a single coordinate list or derives a best-fit transformation 
between two coordinate lists.
&
A single file containing a coordinate list and
a file that describes the transformation or two files, each one is 
containing a coordinate list.
&
A file with the transformed coordinate list in tabulated from or
a file that contains the best-fit transformation.
\\
\texttt{lfit}
&
General purpose arithmetic evaluation, regression and data analysis tool.
&
Files containing data to be analyzed in a tabulated form.
&
Regression parameters or results of the arithmetic evaluation.
\\
\hline
\end{tabularx}
\end{center}
\begin{flushleft}
\end{flushleft}
\end{table*}

This section briefly summarize the main aspects of the implementation
of the package tasks. The tasks can be divided into two
well separated groups with respect to the main purposes. In the first
group there are the programs that manipulate the (astronomical) images
themselves (for instance, read an image, generate one or do a specific 
transformation on a single image or on more images). 
In the second group, there are the tasks that manipulate
textual data, mostly numerical data presented in a tabulated form. \vspace*{1ex}

In general, all of these tasks are capable to the following.
\subsection{Logging and versions}
The codes give release and version information as well as the invocation
can be logged on demand. The version information can be reported by a single
call of the given task, moreover it is logged along with the 
invocation arguments
in the form of special FITS keywords (if the main output of the actual
code is a processed FITS image) and/or in the form of textual comments
(if the main output of the code is text data). Preserving the version 
information along with the invocation arguments makes any kind of output
easily reproducible. 
\subsection{Pipelines}
All of the tasks are capable to read their input from \texttt{stdin} 
(the standard input on UNIX environments) and write the results to 
\texttt{stdout} (the UNIX standard output).
Since many of these tasks manipulate relatively large amount of data,
the number of unnecessary hard disk operations can therefore be 
reduced as small as possible. Moreover, in many cases the output of one 
of the programs is the input of the another one:
pipes, available in all of the modern UNIX-like operating systems,
are basically designed to perform such bindings between the (standard) 
output and (standard) input of two programs. Therefore, such a 
capability of redirecting the input/output data streams significantly 
reduce the overhead of background storage operations.
\subsection{Symbolic operations}
\label{subsec:symbolicoperations}
For the tasks dealing with symbolic operations and
functions, a general back-end 
library\footnote{Available from {http://libpsn.sf.net}, developed by the author.}
is provided to make a user-friendly
interface to specify arithmetic expressions. This kind of approach 
in software systems is barely used, since such a symbolic
specification of arithmetic expressions does not provide a standalone language.
However, it allows an easy and transparent way for arbitrary operations,
and turned out to be very efficient in higher level data reduction scripts.
\subsection{Extensions}
Tasks that manipulate FITS images are capable to 
handle files with multiple extensions. The FITS standard allows the user
to store multiple individual images, as well as (ASCII or binary) tabulated
data in a single file. The control software of some detectors 
produces images that are stored in this extended 
format\footnote{For example, such detectors where the charges from the 
CCD chip are read out in multiple directions: therefore the camera electronics 
utilizes more than one amplifier and A/D converter, thus yield different 
bias and noise levels.}. Other kind of detectors (which acquire 
individual images with a very short exposure time) might store the data in 
the three dimensional format called ``data cube''. The tasks
are also capable to process such data storage formats. 

~

The list of standalone tasks and their main purposes 
that come with the package are shown in Table~\ref{tab:progsummary}. 
The next section discusses these tasks in more details.

\section{Tasks}
\label{sec:tasks}

In this section we summarize the features and properties each of the
standalone tasks that are implemented as distinct binary executables.

\subsection{Basic operations on FITS headers and keywords -- \texttt{fiheader}}

The main purpose of the \texttt{fiheader} utility is to read specific
values from the headers of FITS files and/or alter them on demand. 

Although most of the information about the observational conditions
is stored in the form of FITS keywords, image manipulation programs
use only the necessary ones and most of the image processing
parameters are passed as command line arguments (such keywords and data
are, for example, the gain, the image centroid coordinates, 
astrometrical solutions). The main reasons why this kind of approach
was chosen are the following. 
\begin{list}{\textbullet}{\leftmargin=1.5em \labelsep=0.5em \labelwidth=0.5em \topsep=0ex}
\item First, interpreting many of the standard keywords
leads to false information about the image in the cases of wide-field
or heavily distorted images. Such a parameter is the gain that 
can be highly inhomogeneous for images acquired by an optical system
with non-negligible vignetting and the gain itself cannot be described
by a single real number\footnote{For which the 
\emph{de facto} standard is the \texttt{GAIN} keyword.}, 
rather a polynomial or some equivalent function. Similarly, the standard
World Coordinate System information, describing the astrometrical
solution of the image, has been designed for small field-of-view images,
i.e. the number of coefficients are insufficiently few to properly
constrain the astrometry of a distorted image.
\item Second, altering the meanings of standard keywords leads to 
incompatibilities with existing software. For example, if
the format of the keyword \texttt{GAIN} was changed to be a string of 
finite real numbers (describing a spatially varied gain), other
programs would not be able to parse this redefined keyword.
\end{list}
Therefore, our conclusion was not altering the syntax of the existing keywords,
but 
to define some new (wherever it was necessary). The \texttt{fiheader}
utility enables the user to read any of the keywords, and allows
higher level scripts to interpret the values read from the headers
and pass their values to other programs in the form of command line
arguments.

\subsection{Basic arithmetic operations on images -- \texttt{fiarith}}
\label{sec:prog:fiarith}

The task \texttt{fiarith} allows the user to perform simple
operations on one or more astronomical images. Supposing all of the
input images have the same size, the task allows the user to do
\emph{per pixel} arithmetic operations as well as manipulations 
depend on the pixel coordinates themselves. 
Unlike utilities like \texttt{imarith} in IRAF, \texttt{fiarith} is capable
to process both the symbolic arithmetic operations (done via
the \texttt{libpsn} library, see Sec.~\ref{subsec:symbolicoperations})
and the per-pixel operations via a single call. 

\subsection{Basic information about images -- \texttt{fiinfo}}
\label{sec:prog:fiinfo}

The aim of the task \texttt{fiinfo} is twofold. First, 
this task is capable to gather some statistics and masking 
information of the image. These include
\begin{list}{\textbullet}{\leftmargin=1.5em \labelsep=0.5em \labelwidth=0.5em \topsep=0ex}
\item general statistics, such as mean, median, minimum, maximum,
standard deviation of the pixel values;
\item statistics derived after rejecting the outlier pixels;
\item estimations for the background level and its spatial variations;
\item estimations for the background noise; and
\item the number of masked pixels, detailing for all occurring mask types.
\end{list}
The most common usage of \texttt{fiinfo} in this statistical mode is 
to deselect those calibration frames that seem to be faulty 
(e.g. saturated sky flats, aborted images or so).

Second, the task
is capable to convert astronomical images into widely used 
graphics file formats. Almost all of the
scaling options available in the well known \texttt{DS9} program
\citep[see][]{joye2003} have been implemented in \texttt{fiinfo},
moreover, the user can define arbitrary color palettes as well.
In practice, \texttt{fiinfo} creates only images in PNM (portable 
anymap) format. Images stored in this format can then be converted to any of 
the widely used graphics file formats (such as JPEG, PNG), using existing 
software (e.g. \texttt{netpbm}\footnote{http://netpbm.sourceforge.net/}, 
\texttt{convert}/ImageMagick\footnote{http://www.imagemagick.org/script/index.php}).
Figures in this paper displaying stamps from real (or mock)
astronomical images have also been created using this mode of the task.

\subsection{Combination of images -- \texttt{ficombine}}
\label{sec:prog:ficombine}

The main purpose of image combination is either to create a single image
with from individual images (mainly in order to create higher
signal-to-noise ratio frames and/or images with smaller statistical noise)
or create a mosaic image covering a larger area (mainly from images with
smaller field-of-view). The task \texttt{ficombine} is intended to 
perform this averaging of individual images. This task
has several possible applications in a data
reduction pipeline. For instance, it is used to create the master calibration 
frames (see also Sec.~\ref{sec:prog:ficalib}) or 
the reference frame required by the
method of image subtraction is also created by averaging individual
registered object frames (see also Sec.~\ref{sec:prog:fitrans} about
the details of image registration). 

In the actual implementation, such combination is employed as a 
\emph{per pixel} averaging, where the method of averaging and its
fine tune parameters can be specified via command line arguments. The
most frequently used ``average values'' are the mean and median values.
In many applications, rejection of outlier values are required,
for instance, omitting pixels affected by cosmic ray events. The
respective parameters for tuning the outlier rejection are also
given as command line options. See Sec.~\ref{sec:prog:ficalib} 
for an example about the usage of \texttt{ficombine}, demonstrating
its usage in a simple implementation of a calibration pipeline.

\begin{figure*}
\begin{lcmd}
\#!/bin/sh \\
\hspace*{5mm}~ \\
\# \textit{Names of the individual files storing the raw bias, dark, flat and object frames:} \\
BIASLIST=(\$SOURCE/bias-*.fits) \\
DARKLIST=(\$SOURCE/dark-*.fits) \\
FLATLIST=(\$SOURCE/flat-*.fits) \\
IOBJLIST=(\$SOURCE/target\_object-*.fits) \\
\hspace*{5mm}~ \\
\# \textit{Calibrated images: all the images have the same name but put into a separate directory (\$TARGET):} \\
R\_BIASLIST=(\$(for f in "\$\{BIASLIST[*]\}" ; do echo \$TMP/bias/`basename \${f}` ; done)) \\
R\_DARKLIST=(\$(for f in "\$\{DARKLIST[*]\}" ; do echo \$TMP/dark/`basename \${f}` ; done)) \\
R\_FLATLIST=(\$(for f in "\$\{FLATLIST[*]\}" ; do echo \$TMP/flat/`basename \${f}` ; done)) \\
R\_IOBJLIST=(\$(for f in "\$\{IOBJLIST[*]\}" ; do echo \$TARGET/`basename \${f}` ; done)) \\
\hspace*{5mm}~ \\
\# \textit{Common arguments (saturation level, image section \& trimming, etc.):} \\
COMMON\_ARGS="--saturation 65000 --image 0:0:2047:2047 --trim" \\
\hspace*{5mm}~ \\
\# \textit{The calibration of the individual bias frames, followed by their combination into a single master image:} \\
ficalib -i \$\{BIASLIST[*]\} \$COMMON\_ARGS -o \$\{R\_BIASLIST[*]\}  \\
ficombine \$\{R\_BIASLIST[*]\} --mode median -o \$MASTER/BIAS.fits \\
\hspace*{5mm}~ \\
\# \textit{The calibration of the individual dark frames, followed by their combination into a single master image:} \\
ficalib	-i \$\{DARKLIST[*]\} \$COMMON\_ARGS -o \$\{R\_DARKLIST[*]\} $\backslash$ \\
\ptab	--input-master-bias \$MASTER/BIAS.fits \\
ficombine \$\{R\_DARKLIST[*]\} --mode median -o \$MASTER/DARK.fits \\
\hspace*{5mm}~ \\
\# \textit{The calibration of the individual flat frames, followed by their combination into a single master image:} \\
ficalib	-i \$\{FLATLIST[*]\} \$COMMON\_ARGS --post-scale 20000 -o \$\{R\_FLATLIST[*]\} $\backslash$ \\
\ptab	--input-master-bias \$MASTER/BIAS.fits --input-master-dark \$MASTER/DARK.fits \\
ficombine \$\{R\_FLATLIST[*]\} --mode median -o \$MASTER/FLAT.fits \\
\hspace*{5mm}~ \\
\# \textit{The calibration of the object images:} \\
ficalib	-i \$\{IOBJLIST[*]\} \$COMMON\_ARGS -o \$\{R\_IOBJLIST[*]\} --input-master-bias \$MASTER/BIAS.fits $\backslash$ \\
\ptab	--input-master-dark \$MASTER/DARK.fits --input-master-flat \$MASTER/FLAT.fits
\end{lcmd}
\caption{A shell script demonstrating the proper usage of the 
\texttt{ficalib} and \texttt{ficombine} tasks on the example of a
simple calibration pipeline. The names for the files containing the input 
raw frames (both calibration frames and object frames) are stored in 
the arrays \texttt{\$BIASLIST[*]}, \texttt{\$DARKLIST[*]},
\texttt{\$FLATLIST[*]} and \texttt{\$OBJLIST[*]}. 
In this example, these frames are taken from the directory 
\texttt{\$SOURCE} and their types are identified by 
the file names themselves (hence the wildcard-based selection 
during the declaration of the above arrays).
The variable \texttt{\$COMMON\_ARGS} contains 
all necessary common information related to the frames
(geometry, saturation level, etc.) The individual calibrated bias, dark 
and flat frames are stored in the subdirectories of the 
\texttt{\$TMP} directory. These files are then combined to a single 
master bias, dark and flat frame, that are used in the final step of 
the calibration, when the object frames themselves are calibrated. 
The final calibrated scientific images are stored in the 
directory \texttt{\$TARGET}. Note that each flat frame is scaled after 
calibration to have a mean value of 20,000\,ADU. 
In the case of dome flats, this scaling is not necessary, but in the
case of sky flats, this steps corrects for the variations in the
sky background level (during dusk or dawn). Of course, there are 
may ways to make this simple pipeline to be more sophisticated, depending
on the actual optical and instrumental setup.}
\label{fig:ficalibexample}
\end{figure*}

\subsection{Calibration of images -- \texttt{ficalib}}
\label{sec:prog:ficalib}

In principle, the task \texttt{ficalib} implements 
the evaluation of the equations required by image calibration
in an efficient way. It is optimized for the assumption that
all of the master calibration frames are the same for all of the input images.
Because of this assumption, the calibration process is much more
faster than if it was done independently on each image, using 
the task \texttt{fiarith}.

The task \texttt{ficalib} automatically performs
the overscan correction (if the user specifies overscan regions),
and also trims the image to its designated size (by clipping these 
overscan areas). The output images inherit the masks from
the master calibration images, as well as additional pixels might
be masked from the input images if these were found to be saturated
and/or bloomed. In Fig.~\ref{fig:ficalibexample} a shell script is shown 
that demonstrates the usage of the task \texttt{ficalib} 
and \texttt{ficombine} in brief. 

\subsection{Rejection and masking of nasty pixels -- \texttt{fiign}}
\label{sec:prog:fiign}

The aim of the program \texttt{fiign} is twofold. 
First, it is intended to perform 
low-level operations on masks associated to FITS images, such as
removing some of the masks, converting between layers of the masks 
and merging or combining masks from separate files. Second, various
methods exist with which the user can add additional masks based 
on the image itself. These additional masks can be used to mark 
saturated or blooming pixels, pixels with unexpectedly low and/or high 
values or extremely sharp structures, especially pixels that are 
resulted by cosmic ray events. 

This program is a crucial piece in the calibration pipeline if 
it is implemented using purely the \texttt{fiarith} program. However,
most of the functionality of \texttt{fiign} is also integrated in
\texttt{ficalib} (see Sec.~\ref{sec:prog:ficalib}). Since
\texttt{ficalib} much more efficiently implements the operations
of the calibration than if these were implemented by individual calls 
of \texttt{fiarith}, \texttt{fiign} is used only occasionally in practice.

\subsubsection{Masking}
\label{sec:masking}

As it was mentioned above, pixels having some undesirable properties must be 
masked in order to exclude them from further processing.
The \fitsh{} package and therefore the pipeline of the whole 
reduction supports various kind of masks. These masks are transparently 
stored in the image headers (using special keywords)
and preserved if an independent software modifies the image. 
Technically, this mask is a bit-wise combination of Boolean flags, assigned
to various properties of the pixels. Here we briefly summarize
our masking method.

Actually, the latest version of the \fitsh{} package supports 
the following masks:
\begin{list}{\textbullet}{\leftmargin=1.5em \labelsep=0.5em \labelwidth=0.5em \topsep=0ex}
\item \emph{Mask for faulty pixels.} 
These pixels show strong non-linearity. 
These masks are derived occasionally from the ratios of flat field
images with low and high intensities.
\item \emph{Mask for hot pixels.} The mean dark current for these pixels
is significantly higher than the dark current of normal pixels.
\item \emph{Mask for cosmic rays.} Cosmic rays cause sharp structures,
these structures mostly resemble hot or bad pixels, but these
does not have a fixed structure that is known in advance.
\item \emph{Mask for outer pixels.} After a geometric transformation (dilation,
rotation, registration between two images), certain pixels near the
edges of the frame have no corresponding pixels in the original 
frame. These pixels are masked as ``outer'' pixels. Usually, the pixel 
values assigned to these ``outer'' pixels are zero. 
\item \emph{Mask for oversaturated pixels.} These pixels have an ADU value
that is above a certain limit defined near the maximum value of the A/D
conversion (or below if the detector shows a general nonlinear response
at higher signal levels).
\item \emph{Mask for blooming.} In the cases when the detector has 
no antiblooming feature or this feature is turned off, extremely 
saturated pixels causes ``blooming'' in certain directions (usually 
parallel to the readout direction). The A/D conversion value of the 
blooming pixels does not reach the maximum value of the A/D conversion,
but these pixels also should be treated as somehow saturated. 
The ``blooming'' and ``oversaturated'' pixels are commonly referred
as ``saturated'' pixels, i.e. the logical combination of these two respective
masks indicates pixels that are related to the saturation and its 
side effects.
\item \emph{Mask for interpolated pixels.} Since the cosmic rays and 
hot pixels can be easily detected, in some cases it is worth to 
replace these pixels with an interpolated value derived from the
neighboring pixels. However, these pixels should only be used with
caution, therefore these are indicated by such a mask for the further
processes.
\end{list}
We found that the above categories of 7 distinct masks are feasible
for all kind of applications appearing in the data processing. The fact
that there are 7 masks -- all of which can be stored in a single bit for
a given pixel -- makes the implementation quite easy. All bits of the mask
corresponding to a pixel fit in a byte and we still have an additional bit.
It is rather convenient during the implementation of certain steps (e.g.
the derivation of the blooming mask from the oversaturated mask), since
there is a temporary storage space for a bit that can be used for 
arbitrary purpose.

\begin{figure*}
\begin{center}
\noindent
\resizebox{80mm}{!}{\includegraphics{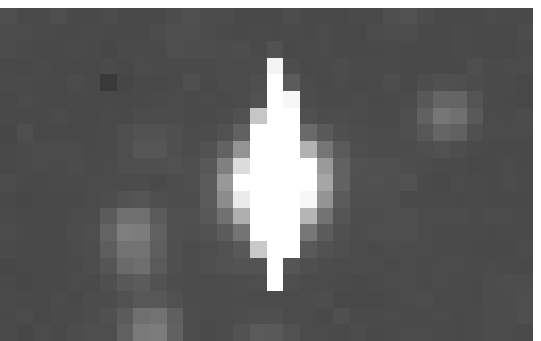}}\hspace*{4mm}%
\resizebox{80mm}{!}{\includegraphics{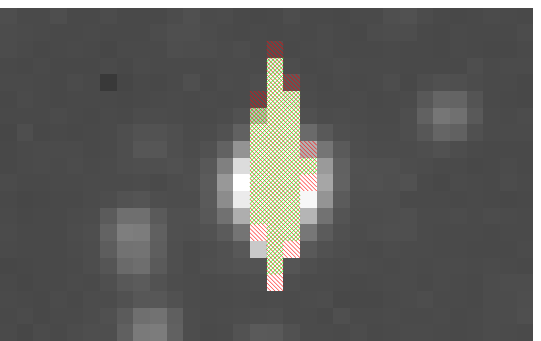}}
\begin{lcmd}
\small
MASKINFO= '1 -32 16,8 -16 0,1:-2 -32 1,1 -2,1 -16 1,0:2 -1,1:3,3 -32 3,2' \\
MASKINFO= '-16 -3,1:4 0,1:3,3 -32 3,0 -3,3 -16 1,0:2 0,1:-2 -32 1,0 -1,2'
\end{lcmd}
\end{center}
\caption{Image stamps showing a typical saturated star and its neighborhood. 
The left panel shows the original image itself where the blooming structure 
can be seen well. The right panel shows the same area where the
oversaturated and bloomed pixels are marked as follows.  
The oversaturated pixels (where the actual ADU values reach the maximum
of the A/D converter) are marked with green right-diagonal stripes while 
pixels affected by blooming are marked with red left-diagonal stripes. 
Note that most of the oversaturated pixels are also marked as blooming 
ones, since their lower and/or upper neighboring pixels are also oversaturated. 
Such pixels are therefore marked with both red and green 
(left- and right-diagonal) stripes. Since the readout direction in 
this particular detector was vertical, the saturation/blooming 
structure is also vertical. The \texttt{``MASKINFO''} blocks seen 
below the two stamps show how this particular masking information
is stored in the FITS headers in a form of special keywords. }
\label{fig:masksaturated}
\end{figure*}

\begin{figure*}
\begin{center}
\footnotesize
\begin{tabularx}{150mm}{lX}
\hline
Value & Interpretation \\
\hline
$T$             &
        Use type $T$ encoding. $T=0$ implies absolute cursor movements,
        $T=1$ implies relative cursor movements. Other values of $T$ are
        reserved for optional further improvements. \\
$-M$            &
        Set the current bitmask to $M$. $M$ must be between 1 and 127 and 
        it is a bit-wise combination of the numbers 1, 2, 4, 8, 16, 
        32 and 64, for faulty, hot, cosmic, outer, oversaturated,
        blooming and interpolated pixels, respectively. \\
$x,y$   &
        Move the cursor to the position $(x,y)$ (in the case of $T=0$)
        or shift the cursor position by $(x,y)$ (in the case of $T=1$)
        and mark the pixel with the mask value of $M$. \\
$x,y:h$         &
        Move/shift the cursor to/by $(x,y)$ and mark the horizontal line
        having the length of $h$ and left endpoint at the actual
        position. \\
$x,y:-v$        &
        Move/shift the cursor to/by $(x,y)$ and mark the vertical line 
        having the length of $v$ and lower endpoint at the actual
        position. \\
$x,y:h,w$       &
        Move/shift the cursor to/by $(x,y)$ and mark the rectangle
        having a size of $h\times w$ and lower-left corner at the actual
        cursor position. \\
\hline
\end{tabularx}
\end{center}
\caption{Interpretation of the tags found 
\texttt{MASKINFO} keywords in order to decode the respective mask. The
values of $M$, $h$, $v$ and $w$ must be always positive.}
\label{fig:maskalgorithm}
\end{figure*}

\begin{figure*}
\begin{center}
\resizebox{50mm}{!}{\includegraphics{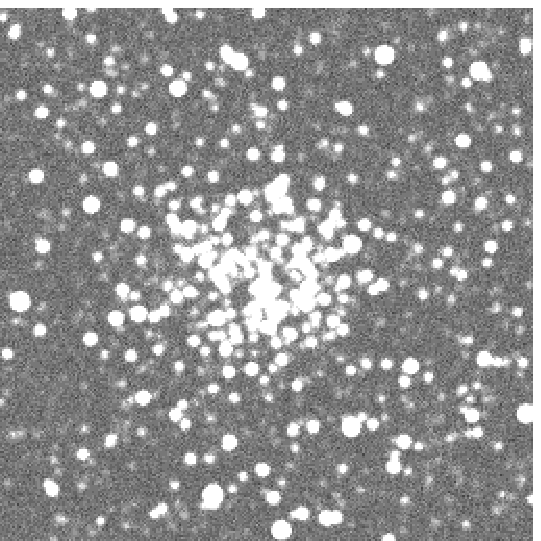}}\hspace*{5mm}%
\resizebox{50mm}{!}{\includegraphics{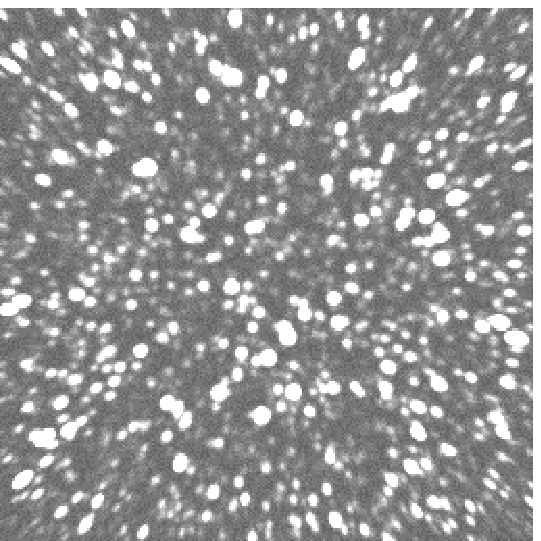}}\hspace*{5mm}%
\resizebox{50mm}{!}{\includegraphics{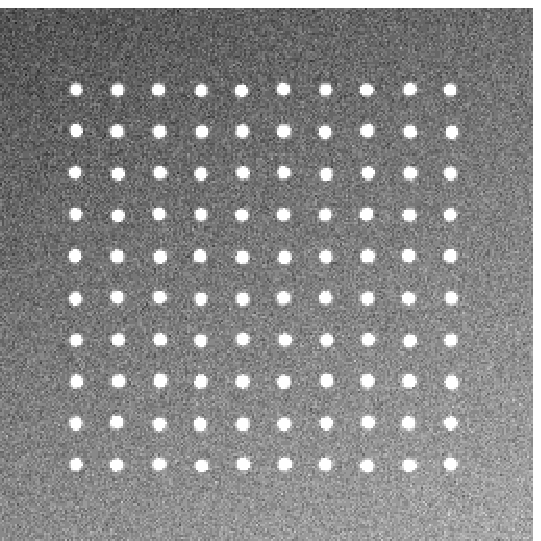}}
\begin{lcmd}
\#!/bin/sh \\
\hspace*{5mm}~ \\
firandom --size 256,256 $\backslash$ \\
\ptab	~--list "f=3.2,500*[x=g(0,0.2),y=g(0,0.2),m=15-5*r(0,1)\^{ }2]" $\backslash$ \\
\ptab	~--list "f=3.2,1400*[x=r(-1,1),y=r(-1,1),m=15+1.38*log(r(0,1))]" $\backslash$ \\
\ptab	~--sky 100 --sky-noise 10 --integral --photon-noise --bitpix -32 --output globular.fits \\
\hspace*{5mm}~ \\
firandom --size 256,256 $\backslash$ \\
\ptab	~--list "5000*[x=r(-1,1),y=r(-1,1),s=1.3,d=0.3*(x*x-y*y),k=0.6*x*y,m=15+1.38*log(r(0,1))]" $\backslash$ \\
\ptab	~--sky 100 --sky-noise 10 --integral --photon-noise --bitpix -32 --output coma.fits \\
\hspace*{5mm}~ \\
firandom --size 256,256 $\backslash$ \\
\ptab	~--list "f=3.0,100*[X=36+20*div(n,10)+r(0,1),Y=36+20*mod(n,10)+r(0,1),m=10]" $\backslash$ \\
\ptab	~--sky "100+x*10-y*20" --sky-noise 10 --integral --photon-noise --bitpix -32 --output grid.fits \\
\hspace*{5mm}~ \\
for base in globular coma grid ; do \\
\ptab	fiinfo	 \$\{base\}.fits --pgm linear,zscale --output-pgm - | pnmtoeps -g -4 -d -o \$\{base\}.eps  \\
done
\end{lcmd}
\end{center}
\caption{Three mock images generated using the program
\texttt{firandom}. The first image (\texttt{globular.fits}) on the left 
shows a ``globular cluster'' with some field stars as well. For simplicity,
the distribution of the cluster stars are Gaussian and the magnitude
distribution is quadratic while the field stars distribute uniformly and
their magnitudes is derived from assuming uniformly distributed stars of 
constant brightness. The second image (\texttt{coma.fits}) simulates
nearly similar effect on the stellar profiles what comatic aberration 
would cause. The shape parameters $\delta$ and $\kappa$ (referred 
as \texttt{d} and \texttt{k} in the command line argument of the program)
are specific functions of the spatial coordinates. The magnitude 
distribution of the stars is the same as for the field stars in
the previous image. The third image (\texttt{grid.fits})
shows a set of stars positioned
on a grid. The background of this image is not constant.
The shell script below the image stamps is used to create these FITS files.
The body of the last iterator loop in the script converts the FITS files 
into PGM format, using the \texttt{fiinfo} utility (see 
Sec.~\ref{sec:prog:fiinfo}) and the well-known 
\texttt{zscale} intensity scaling algorithm \citep[see DS9,][]{joye2003}.
The images yielded by \texttt{fiinfo} are instantly converted to EPS 
(encapsulated Postscript) files, that is the preferred format for many
typesetting systems, such as \LaTeX.}
\label{fig:firandomexamples}
\end{figure*}

\subsection{Generation of artificial images -- \texttt{firandom}}
\label{sec:prog:firandom}

The main purpose of the program \texttt{firandom} is to create
artificial images. These artificial images can be used either to 
create \emph{model images} for real observations 
(for instance, to remove fitted stellar PSFS) or \emph{mock images}
that are intended to simulate some of the influence related to
one or more observational artifacts and realistic effects. 
In principle, \texttt{firandom}
creates an image with a given background level on which sources are drawn.
Additionally, \texttt{firandom} is capable to add noise to the images,
simulating both the effect of readout and background noise as well
as photon noise. In the case of mock images, \texttt{firandom} is 
also capable to generate the object list itself. Moreover, \texttt{firandom}
is capable to draw stellar profiles derived from PSFs (by
the program \texttt{fistar}, see also Sec.~\ref{sec:prog:fistar}).

The program features symbolic input processing, i.e. the variations
in the background level, the spatial distribution of the object 
centroids (in the case of mock images), the profile shape parameters,
fluxes for individual objects and the noise level can be specified
not only as a tabulated dataset but in the form of arithmetic
expressions. In these expressions one can involve various built-in 
arithmetic operators and functions, including random number generators.
Of course, the generated mock coordinate lists can also be saved
in tabulated form. In Fig.~\ref{fig:firandomexamples}, 
some examples are shown that demonstrate the usage of the 
program \texttt{firandom}. 

\subsection{Detection of stars or point-like sources -- \texttt{fistar}}
\label{sec:prog:fistar}

The source detection and stellar profile modelling algorithms 
\citep[see][]{pal2009phd}
are implemented in the program \texttt{fistar}. The main purpose of this
program is therefore to search for and characterize point-like sources.
Additionally, the program is capable to derive the point-spread function
of the image, and spatial variations of the PSF can also be fitted
up to arbitrary polynomial order. 

The list of detected sources, their centroid coordinates, shape
parameters (including FWHM) and flux estimations are written to a
previously defined output file. This file can have arbitrary format,
depending on our needs. The best fit PSF is saved in FITS format. If
the PSF is supposed to be constant throughout the image, 
the FITS image is a normal two-dimensional image. Otherwise,
the PSF data and the associated polynomial coefficients are stored
in ``data cube'' format, and the size of the $z$ (\texttt{NAXIS3}) axis
is $(N_{\rm PSF}+1)(N_{\rm PSF}+2)/2$, where $N_{\rm PSF}$
is the polynomial order used for fitting the spatial variations.

\subsection{Basic coordinate list manipulations -- \texttt{grtrans}}
\label{sec:prog:grtrans}

The main purpose of the program \texttt{grtrans} is to perform
coordinate list transformations, mostly related to stellar profile
centroid coordinates and astrometrical transformations. 
Since this program is used exhaustively with the program
\texttt{grmatch}, examples and further discussion of this
program can be found in the next section, Sec.~\ref{sec:prog:grmatch}.

\subsection{Matching lists or catalogues -- \texttt{grmatch}}
\label{sec:prog:grmatch}

The main purpose of the \texttt{grmatch} code is to implement 
the point matching algorithm that is the key point in the derivation
of the astrometric solution and source identification. See 
\cite{pal2006} or \cite{pal2009phd} about more details on the algorithm 
itself. We note here that although the program \texttt{grmatch} is 
sufficient for point matching and source identification purposes,
one may need other codes to conveniently 
interpret or use the output of this program.
For instance, tabulated list of coordinates can be transformed 
from one reference frame to another, using the program \texttt{grtrans} 
while the program \texttt{fitrans} is capable to apply these derived 
transformations on FITS images, in order to, e.g. register 
images to the same reference frame.

\begin{figure*}
\begin{lcmd}
\ptab \\
for base in \$\{LIST\_OF\_FRAMES[*]\} ; do \\
\ptab \\
\ptab	grmatch --reference \$CATALOG --col-ref \$COL\_X,\$COL\_Y --col-ref-ordering -\$COL\_MAG $\backslash$ \\
\ptab	\ptab	--input \$AST/\$base.stars --col-inp 2,3 --col-inp-ordering +8 $\backslash$ \\
\ptab	\ptab	--weight reference,column=\$COL\_MAG,magnitude,power=2 $\backslash$ \\
\ptab	\ptab	--order \$AST\_ORDER --max-distance \$MAX\_DISTANCE $\backslash$ \\
\ptab	\ptab	--output-transformation \$AST/\$base.trans --output \$AST/\$base.match || break \\
\ptab	\ptab \\
\ptab   grtrans \$CATALOG $\backslash$ \\
\ptab	\ptab	--col-xy  \$COL\_X,\$COL\_Y --input-transformation \$AST/\$base.trans $\backslash$ \\
\ptab	\ptab	--col-out \$COL\_X,\$COL\_Y --output - | $\backslash$ \\
\ptab	grmatch --reference - --col-ref \$COL\_X,\$COL\_Y --input \$AST/\$base.stars --col-inp 2,3 $\backslash$ \\
\ptab	\ptab	--match-coords --max-distance \$MAX\_MATCHDST --output - | $\backslash$ \\
\ptab	grtrans --col-xy  \$COL\_X,\$COL\_Y --input-transformation \$AST/\$base.trans --reverse $\backslash$ \\
\ptab	\ptab	--col-out \$COL\_X,\$COL\_Y --output \$AST/\$base.match \\
\ptab \\
done \\
\ptab 
\end{lcmd}
\caption{A typical application for the 
\texttt{grmatch} -- \texttt{grtrans} programs, for the cases where a few of
the stars have high proper motion thus have significant offsets from
the catalogue positions. For each frame (named \texttt{\$base}), the input 
catalogue (\texttt{\$CATALOG}) is matched with the respective list
of extracted stars (found in the \texttt{\$AST/\$base.stars} file), keeping a
relatively large maximum distance between the nominal and detected stellar
positions (\texttt{\$MAX\_DISTANCE}, e.g. $4-6$ pixels, derived from the
expected magnitude of the proper motions from the catalogue epoch and the
approximate plate scale). This first initial match identifies all of the
sources (including the ones with large proper motion),
stored in \texttt{\$AST/\$base.match} file in the form of matched
detected source and catalogue entries. However, the astrometric 
transformation (stored in \texttt{\$AST/\$base.trans}) is systematically 
affected by these high proper motion stars. In order to get rid of this
effect, the match is performed again by excluding the stars 
with higher residual distance (by setting \texttt{\$MAX\_MACHDIST} to
e.g. $1-2$ pixels). 
The procedure is then repeated for all frames (elements of the 
\texttt{\$LIST\_OF\_FRAMES[]} array) in the similar manner.
}
\label{fig:grmatchgrtransexample}
\end{figure*}

\subsubsection{Typical applications}
\label{sec:grmatch:applications}

As it was mentioned earlier, the programs \texttt{grmatch} and \texttt{grtrans}
are generally involved in a complete photometry pipeline right 
after the star detection and before the instrumental photometry. 
If the accuracy of the coordinates in the 
reference catalogue is sufficient to yield a consistent plate solution,
one can obtain the photometric centroids by simply invoking 
these programs. A more sophisticated example for these program is 
shown in Fig.~\ref{fig:grmatchgrtransexample}. In this example these programs 
are invoked twice in order to both derive a proper astrometric 
solution\footnote{By taking into account only the stars with negligible
proper motion.} and properly identify the stars with larger intrinsic proper 
motion\footnote{That would otherwise significantly distort the 
astrometric solution.}. A simple direct application of \texttt{grmatch} and 
\texttt{grtrans} as a part of a complete photometric pipeline 
is displayed in Fig.~\ref{fig:apphotexample}.

\subsection{Transforming and registering images -- \texttt{fitrans}}
\label{sec:prog:fitrans}

As it is known \citep[Sec.~2.8]{pal2009phd}, the 
image convolution and subtraction process requires the images
to be in the same spatial reference system. The details of this
registration process and the underlying algorithms have been explained in 
Sec.~2.7 of \cite{pal2009phd}. The purpose of the program 
\texttt{fitrans} is to implement these various image interpolation
methods.

In principle, \texttt{fitrans} reads an image and a transformation
file, performs the spatial transformation and writes the output image
to a separate file. Image data are read from FITS files while
the transformation files are presumably derived from the appropriate 
astrometric solutions. The output of the \texttt{grmatch} and 
\texttt{grtrans} programs can be directly passed to \texttt{fitrans}.
Of course, \texttt{fitrans} takes into account the masks associated
to the given image as well as derive the appropriate mask for the
output file. Pixels which cannot be mapped from the original image
have always a value of zero and these are marked
as \emph{outer pixels} (see also Sec.~\ref{sec:masking}).
Modern imaging systems are deployed with high-resolution detectors,
therefore the spatial transformation involving exact integration on
biquadratic interpolation surfaces might be a computationally expensive
process \citep[see][Sec.~2.6.3]{pal2009phd}. However, distinct image 
transformations can be performed independently (i.e. a given transformation 
does not have any influence on another transformations), thus the 
complete registration process can easily be performed in parallel. 

\subsection{Convolution and image subtraction -- \texttt{ficonv}}
\label{sec:prog:ficonv}

This member of the \fitsh{} package is intended to implement the
tasks related to the kernel fit, image convolution and subtraction. 
In principle, \texttt{ficonv} has two basic modes. First, assuming an existing 
kernel solution, it evaluates the basic convolution equations
\citep[Eq.~67]{pal2009phd} on 
an image and writes the convolved result to a separate image file.
Second, assuming a base set of kernel functions 
\citep[Eq.~73]{pal2009phd} and some model for the 
background variations \citep[Eq.~75]{pal2009phd}
it derives the best fit kernel solution for the basic convolution equations.
Since this fit yields a linear equation for these coefficients, the
method of classic linear least squares minimization can be efficiently
applied. However, the least squares matrix can have a relatively
large dimension in the cases where the kernel basis is also large and/or
higher order spatial variations are allowed. In the fit mode,
the program yields the kernel solution, and optionally
the convolved and the subtracted residual image 
can also be saved into separate files
without additional invocations of \texttt{ficonv}
and/or \texttt{fiarith}.

The program \texttt{ficonv} also implements the fit for cross-convolution
kernels \citep[Eq.~79]{pal2009phd}. In 
this case, the two kernel solutions are saved to two distinct files.
Subsequent invocations of \texttt{ficonv} and/or \texttt{fiarith} 
can then be used to analyze various kinds of outputs.

Sec.~2.9 of \cite{pal2009phd} discusses the
relevance of the kernel solution in the case when the photometry
is performed on the residual (subtracted) images. The best fit 
kernel solution obtained by \texttt{ficonv} has to be directly passed
to the program \texttt{fiphot} (Sec.~\ref{sec:prog:fiphot}) in order
to properly take into account the convolution information during the 
photometry \citep[Eq.~83]{pal2009phd}.

\begin{figure*}
\begin{lcmd}
SELF=\$0; base="\$1" \\
if [ -n "\$base" ] ; then \\
\ptab	fitrans	\$\{FITS\}/\$base.fits $\backslash$ \\
\ptab	\ptab	--input-transformation \$\{AST\}/\$base.trans --reverse -k -o \$\{REG\}/\$base-trans.fits  \\
else \\
\ptab	pexec -f BASE.list -e base -o - -u - -c -- "\$SELF $\backslash$\$base" \\
fi
\end{lcmd}
\begin{lcmd}
SELF=\$0; base="\$1" \\
if [ -n "\$base" ] ; then \\
\ptab	KERNEL="i/4;b/4;d=3/4" \\
\ptab	ficonv ~--reference ./photref.fits $\backslash$ \\
\ptab	\ptab	--input \$\{REG\}/\$base-trans.fits --input-stamps ./photref.reg --kernel "\$KERNEL" $\backslash$ \\
\ptab	\ptab	--output-kernel-list \$\{AST\}/\$base.kernel --output-subtracted \$\{REG\}/\$base-sub.fits \\
else \\
\ptab	pexec -f BASE.list -e base -o - -u - -c -- "\$SELF $\backslash$\$base" \\
fi
\end{lcmd}
\caption{Two shell scripts demonstrating the invocation syntax of
the \texttt{fitrans} and \texttt{ficonv}. Since the computation of
the transformed and convolved images require significant amount of CPU
time, the utility \texttt{pexec} (\texttt{http://www.gnu.org/software/pexec})
is used to run the jobs in parallel on multiple CPUs.}
\label{fig:imgsubexample}
\end{figure*}

\subsection{Photometry -- \texttt{fiphot}}
\label{sec:prog:fiphot}

The program \texttt{fiphot} is the main code in the \fitsh{} package
that performs the raw and instrumental photometry. In the
current implementation, we were focusing on the aperture photometry,
performed on normal and subtracted images. Basically, \texttt{fiphot}
reads an astronomical image (FITS file) and a centroid list file, 
where the latter
should contain not only the centroid coordinates but the individual 
object identifiers as well\footnote{If the proper object identification
is omitted, \texttt{fiphot} assigns some arbitrary (but indeed unique)
identifiers to the centroids, however, in practice it is almost useless.}.

In case of image subtraction-based photometry, \texttt{fiphot}
requires also the kernel solution (derived by \texttt{ficonv}). Otherwise,
if this information is omitted, the results of the photometry are not
reliable and consistent \citep[Sec.~2.9]{pal2009phd}.

In Fig.~\ref{fig:apphotexample}, a complete shell script
is displayed, as an example of various \fitsh{} programs related
to the photometry process.

Currently, PSF photometry is not implemented directly in the 
program \texttt{fiphot}. However, the program \texttt{fistar} 
(Sec.~\ref{sec:prog:fistar}) is capable to do PSF fitting on 
the detected centroids, although its output is not compatible
with that of \texttt{fiphot}. Alternatively, \texttt{lfit}
(see Sec.~\ref{sec:prog:lfit}) can be used to perform profile fitting,
if the pixel intensities are converted to ASCII tables in 
advance\footnote{The program
\texttt{fiinfo} is capable to produce such tables with three columns:
a list of $x$ and $y$ coordinates followed by the respective pixel 
intensity.}, however, it is not computationally efficient. 

\begin{figure*}
\begin{center}
\begin{ccmd}
\textit{\# \$\{PHOT\}/IMG-1.phot:} \\
IMG-1 STAR-01 6.8765 0.0012 C \\
IMG-1 STAR-02 7.1245 0.0019 G \\
IMG-1 STAR-03 7.5645 0.0022 G \\
IMG-1 STAR-04 8.3381 0.0028 G \\
\hspace*{5mm}~ \\
\hspace*{5mm}~ \\
\textit{\# \$\{PHOT\}/IMG-2.phot:} \\
IMG-2 STAR-01 6.8778 0.0012 C \\
IMG-2 STAR-02 7.1245 0.0020 G \\
IMG-2 STAR-03 7.5657 0.0023 G \\
IMG-2 STAR-04 8.3399 0.0029 G \\
\hspace*{5mm}~ \\
\hspace*{5mm}~ \\
\textit{\# \$\{PHOT\}/IMG-3.phot:} \\
IMG-3 STAR-01 6.8753 0.0012 G \\
IMG-3 STAR-02 7.1269 0.0019 G \\
IMG-3 STAR-03 7.5652 0.0023 G \\
IMG-3 STAR-04 8.3377 0.0029 G 
\end{ccmd}\hspace*{3.20ex}$\Rightarrow$\hspace*{3.20ex}%
\begin{ccmd}
\textit{\# \$\{LC\}/STAR-01.lc:} \\
IMG-1 STAR-01 6.8765 0.0012 C \\
IMG-2 STAR-01 6.8778 0.0012 C \\
IMG-3 STAR-01 6.8753 0.0012 G \\
\hspace*{5mm}~ \\
\textit{\# \$\{LC\}/STAR-02.lc:} \\
IMG-1 STAR-02 7.1245 0.0019 G \\
IMG-2 STAR-02 7.1245 0.0020 G \\
IMG-3 STAR-02 7.1269 0.0019 G \\
\hspace*{5mm}~ \\
\textit{\# \$\{LC\}/STAR-03.lc:} \\
IMG-1 STAR-03 7.5645 0.0022 G \\
IMG-2 STAR-03 7.5657 0.0023 G \\
IMG-3 STAR-03 7.5652 0.0023 G \\
\hspace*{5mm}~ \\
\textit{\# \$\{LC\}/STAR-04.lc:} \\
IMG-1 STAR-04 8.3381 0.0028 G \\
IMG-2 STAR-04 8.3399 0.0029 G \\
IMG-3 STAR-04 8.3377 0.0029 G   
\end{ccmd}
\begin{lcmd}
grcollect \$\{PHOT\}/IMG-*.phot --col-base 2 --prefix \$\{LC\}/ --extension lc --max-memory 256m \\
\hspace*{5mm}~ \\
cat \$\{PHOT\}/IMG-*.phot | grcollect - --col-base 2 --prefix \$\{LC\}/ --extension lc --max-memory~256m
\end{lcmd}
\end{center}
\caption{The schematics of the data transposition. Records for
individual measurements are written initially to photometry files
(having an extension of \texttt{*.phot}, for instance). These records
contain the source identifiers. During the transposition, photometry 
files are converted to light curves. In principle, these light curves
contain the same records but sorted into distinct files by 
the object names, not the frame identifiers. The command lines on
the lower panel show some examples how this data transposition
can be employed involving the program \texttt{grcollect}.}
\label{fig:transpositionbasics}
\end{figure*}

\subsection{Transposition of tabulated data -- \texttt{grcollect}}
\label{sec:prog:grcollect}

Raw and instrumental photometric data obtained for each frame are 
stored in separate files by default as it was discussed earlier 
(see also Sec.~\ref{sec:prog:fiphot}).
We refer to these files as \emph{photometric files}. In order to analyze 
the per-object outcome of our data reductions, one has to have
the data in the form of \emph{light curve files}. Therefore,
the step of photometry (including the magnitude transformation)
is followed immediately by the step of \emph{transposition}. See
Fig.~\ref{fig:transpositionbasics} about how this step looks like in
a simple case of 3 photometric files and 4 objects. 

The main purpose of the program \texttt{grcollect} is to perform
this transposition on the photometric data in order to have 
the measurements being stored in the form of light curves and therefore
to be adequate for further per-object analysis (such as 
light curve modelling). The invocation syntax
of \texttt{grcollect} is also shown in Fig.~\ref{fig:transpositionbasics}.
Basically, small amount of information is needed 
for the transposition process: the name of the input files, the 
index of the column in which the object identifiers are stored and the optional
prefixes and/or suffixes for the individual light curve file names.
The maximum memory that the program is allowed to use is also specified
in the command line argument. In fact, \texttt{grcollect} does not 
need the original data to be stored in separate files. The second example
on Fig.~\ref{fig:transpositionbasics} shows an alternate
way of performing the transposition, namely 
when the whole data is read from the standard input (and the 
preceding command of \texttt{cat} dumps all the data to the standard output,
these two commands are connected by a single uni-directional pipe).

\begin{figure}
\begin{center}
\noindent
\resizebox{80mm}{!}{\includegraphics{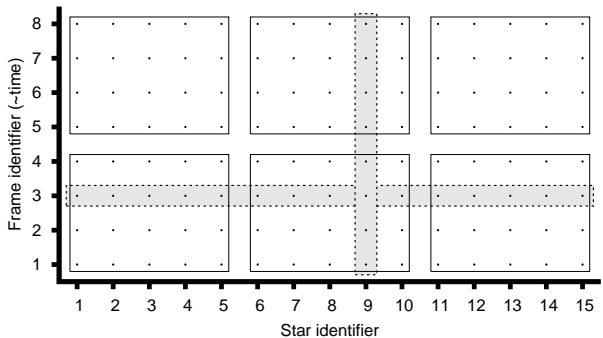}}
\end{center}\vspace*{-5mm}
\caption{Storage schemes for photometric data. Supposing a series of 
frames, on which nearly the same set of stars have individual photometric
measurements, the figure shows how these data can be arranged 
for practical usages. The target stars (their identifiers) are 
arranged along the abscissa while the ordinate shows the frame identifiers
to which individual measurements (symbolized by dots) belong.
Raw and instrumental photometric data are therefore represented here
as rows (see the marked horizontal stripe for frame \#3, for instance) while 
the columns refer to light curves. In practice, native ways of transposition 
are extremely ineffective if the total amount of data does not fit into
the memory. The transposition can be speeded up by using an intermediate
stage of data storage, so-called macroblocks. In the figure, each macroblock
is marked by an enclosing rectangle. See text for further details. }
\label{fig:macroblock}
\end{figure}

The actual implementation of the transposition inside \texttt{grcollect}
is very simple: it reads the data from the individual files (or from
the standard input) until the data fit in the available memory. 
If this temporary memory is full of records, this array is sorted by 
the object identifier and 
the sorted records are written/concatenated to distinct files.
The output files are named based on the appropriate object identifiers. 
This procedure is repeated until there are available data. 
Although this method creates the light curve files, it means
that neither the whole process nor the access 
to these light curve files is effective. If either the number of
the files to be transposed or the number of the records in a single
file exceed a certain limit (that can be derived from the
available memory, the record size and the ratio of the disk average
seek time and sequential input/output bandwidth), the 
whole transposition process becomes highly ineffective. 
This can be avoided by introducing some intermediate stages of transpositions,
see Fig.~\ref{fig:macroblock} or \cite{pal2009phd} for further details.

\begin{table*}
\caption{Algorithms supported 
by \texttt{lfit} and their respective 
requirements for the model function. The first column refers
to the internal and command line identifier of the algorithms. The second
column shows whether the method requires the parametric derivatives of the 
model functions in an analytic form or not. The third column indicates
whether in the cases when the method requires parametric derivatives, 
should the model function be linear in all of the parameters.
}\label{table:lfitmethods}
\begin{center}
\footnotesize
\begin{tabularx}{160mm}{lllX}
\hline
Code		& derivatives		& linearity		& Method or algorithm	\\
\hline
\texttt{L/CLLS}	& yes			& yes			& Classic linear least squares method	\\
\texttt{N/NLLM}	& yes			& no			& (Nonlinear) Levenberg-Marquardt algorithm \\
\texttt{U/LMND}	& no			& no			& Levenberg-Marquardt algorithm employing numeric parametric derivatives \\
\texttt{M/MCMC}	& no			& no			& Classic Markov Chain Monte-Carlo algorithm$^{1}$	\\
\texttt{X/XMMC}	& yes			& no			& Extended Markov Chain Monte-Carlo$^{2}$ \\
\texttt{K/MCHI}	& no			& no			& Mapping the values $\chi^2$ on a grid	(a.k.a. ``brute force'' minimization) \\
\texttt{D/DHSX}	& optional$^{3}$	& no			& Downhill simplex \\
\texttt{E/EMCE}	& optional$^{4}$	& optional$^{4}$	& Uncertainties estimated by refitting to synthetic data sets	\\
\texttt{A/FIMA}	& yes			& no			& Fisher Information Matrix Analysis \\
\hline
\end{tabularx}
\end{center}
\begin{minipage}{\textwidth}
\scriptsize
\noindent \hspace*{1ex} $^{1}$
	The implemented transition function is based on the
	Metropolitan-Hastings algorithm and the optional Gibbs sampler.
	The transition amplitudes must be specified initially. Iterative
	MCMC can be implemented by subsequent calls of \texttt{lfit}, involving
	the previous inverse statistical variances for each parameters as 
	the transition amplitudes for the next chain.

\noindent \hspace*{1ex} $^{2}$
	The also program reports the summary related to the sanity 
	checks (such as correlation lengths, Fisher covariance, statistical
	covariance, transition probabilities and the best fit value obtained
	by an alternate /usually the downhill simplex/ minimization).

\noindent \hspace*{1ex} $^{3}$
	The downhill simplex algorithm may use the parametric derivatives to
	estimate the Fisher/covariance matrix for the initial conditions
	in order to define the control points of the initial simplex.
	Otherwise, if the parametric derivatives do not exist, the 
	user should specify the ``size'' of the initial simplex somehow
	in during the invocation of \texttt{lfit}.

\noindent \hspace*{1ex} $^{4}$
	Some of the other methods (esp. CLLS, NLLM, DHSX, in practice)
	can be used during the minimization process of the original data
	and the individual synthetic data sets.
\end{minipage}
\end{table*}

\subsection{Archiving -- \texttt{fizip} and \texttt{fiunzip}}
\label{sec:prog:fizip}

Due to the large disk space required to store the raw, calibrated
and the derived (registered and/or subtracted) frames,
it is essential to compress and archive the image files that are barely used.
The purpose of the \prog{fizip} and \prog{fiunzip} programs is to compress and 
decompress primary FITS data, by keeping the changes in the primary FITS
header to be minimal. The compressed data is stored in a one-dimensional
8 bit (\texttt{BITPIX=8}, \texttt{NAXIS=1}) array, therefore these 
keywords does not reflect the original image dimension or data type. 

All of the other keywords
are untouched. Some auxiliary information on the compression is stored in
the keywords starting with ``\texttt{FIZIP}'', the contents of these keywords
depend on the involved compression method. \prog{fizip} rejects compressing
FITS file where such keywords exist in the primary header.

In practice, \prog{fizip} and \prog{fiunzip} refer to the same
program (namely, \prog{fiunzip} is a symbolic link to \prog{fizip})
since the algorithms involved in the compression and decompression 
refer to the same codebase or external library. \prog{fizip} and \prog{fiunzip}
support well known compression algorithms, such as the GNU zip (``gzip'')
and the block-sorting file compressor (also known as ``bzip2'') algorithm.

These compression algorithms are lossless. However, \prog{fizip}
supports rounding the input pixel values to the nearest integer 
or to the nearest fraction of some power of 2. Since the common representation
of floating-point real numbers yields many zero bits if the number itself
is an integer or a multiple of power of 2 (including fractional 
multiples), the compression is more effective if this kind of 
rounding is done before the compression. This ``fractional rounding''
yields data loss. However, if the difference between the original and 
the rounded values are comparable or less than the readout noise
of the detector, such compression does not affect the quality of the further
processing (e.g. photometry).

\subsection{Generic arithmetic evaluation, regression and data analysis -- \texttt{lfit}}
\label{sec:prog:lfit}

Modeling of data is a prominent step in the analysis 
and interpretation of astronomical observations. In 
this section, a standalone command line driven tool, 
named \texttt{lfit} is introduced, designed for both interactive 
and batch processed regression analysis as well as generic 
arithmetic evaluation. 

Similarly to the task \texttt{fiarith}, this tool is built on the top of 
the \texttt{libpsn} library (see Sec.~\ref{subsec:symbolicoperations}). 
This library provides 
both the back-end for function evaluation as well as analytical calculations
of partial derivatives. Partial derivatives are required by most of 
the regression methods 
(e.g. linear and non-linear least squares fitting) and uncertainty 
estimations (e.g. Fisher analysis). The program features many built-in
functions related to special astrophysical problems. Moreover, it allows
the end-user to extend the capabilities during run-time using dynamically
loaded libraries. Sophisticated functions can be implemented and 
passed to \texttt{lfit} via 
this way \citep[see e.g.][for a practical case]{pal2010}.

The built-in regression methods (Table~\ref{table:lfitmethods}), 
the built-in functions related
to astronomical data analysis (see also Table~\ref{table:lfitastromext})
and some of the more sophisticated 
new tools for nonlinear regression analyses (e.g. extended 
Markov Chain Monte-Carlo, XMMC) are discussed in \cite{pal2009phd}
in more details. 

\begin{table*}
\caption{Basic functions found in the built-in astronomical 
extension library. These
functions cover the fields of simple radial velocity analysis,
some aspects of light curve modelling and data reduction. These functions
are a kind of ``common denominators'', i.e. they do not provide a direct
possibility for applications but complex functions can be built on the
top of them for any particular usage. All of the functions below
with the exception of \texttt{hjd()} and \texttt{bjd()} 
have partial derivatives that can be evaluated 
analytically by \texttt{lfit}.}
\label{table:lfitastromext}
\begin{center}
\footnotesize
\begin{tabularx}{\textwidth}{lX}
\hline
Function	& Description	\\
\hline
${\tt hjd}({\rm JD},\alpha,\delta)$	&
	Function that calculates the 
	heliocentric Julian date from the Julian day $J$ and
	the celestial coordinates $\alpha$ (right ascension)
	and $\delta$ (declination). \\

${\tt bjd}({\rm JD},\alpha,\delta)$	&
	Function that calculates the 
	barycentric Julian date from the Julian day $J$ and
	the celestial coordinates $\alpha$ (right ascension)
	and $\delta$ (declination). \\
${\tt ellipticK}(k)$		&
	Complete elliptic integral of the first kind.	 \\
${\tt ellipticE}(k)$		&
	Complete elliptic integral of the second kind.	 \\
${\tt ellipticPi}(k,n)$		&
	Complete elliptic integral of the third kind.	 \\
${\tt eoq}(\lambda,k,h)$	&
	Eccentric offset function, `q` component. The arguments
	are the mean longitude $\lambda$, in radians and the 
	Lagrangian orbital elements $k=e\cos\varpi$, $h=e\sin\varpi$. \\
${\tt eop}(\lambda,k,h)$	&
	Eccentric offset function, `p` component. \\
${\tt ntiu}(p,z)$		&
	Normalized occultation flux decrease. This function calculates the
	flux decrease during the eclipse of two spheres when one of the
	spheres has uniform flux distribution and the other one by which
	the former is eclipsed is totally dark. The bright source is assumed
	to have a unity radius while the occulting disk has a radius of $p$.
	The distance between the centers of the two disks is $z$.	\\
${\tt ntiq}(p,z,\gamma_1,\gamma_2)$		&
	Normalized occultation flux decrease when eclipsed sphere has a 
	non-uniform flux distribution modelled by quadratic limb darkening law. 
	The limb darkening is characterized 
	by $\gamma_1$ and $\gamma_2$. \\
\hline
\end{tabularx}
\end{center}
\end{table*}

\begin{figure*}
\begin{lcmd}
	\# This command just prints the content of the file ``line.dat'' to the standard output: \\
	\$ \textit{cat line.dat} \\
	2 ~8.10 \\
	3 10.90 \\
	4 14.05 \\
	5 16.95 \\
	6 19.90 \\
	7 23.10 \\
	\# Regression: this command fits a ``straight line'' to the above data: \\
	\$ \textit{lfit -c x,y -v a,b -f "a*x+b" -y y line.dat} \\
	\hspace*{0ex}~~~~~2.99714~~~~~~2.01286 \\
	\# Evaluation: this command evaluates the model function assuming the parameters to be known: \\
	\$ \textit{lfit -c x,y -v a=2.99714,b=2.01286 -f "x,y,a*x+b,y-(a*x+b)" -F \%6.4g,\%8.2f,\%8.4f,\%8.4f line.dat} \\
	\hspace*{0ex}~~~~~  2~~~~~~~~~8.10~~~~~~8.0071~~~~~~0.0929 \\
	\hspace*{0ex}~~~~~  3~~~~~~~~10.90~~~~~11.0043~~~~~-0.1042 \\
	\hspace*{0ex}~~~~~  4~~~~~~~~14.05~~~~~14.0014~~~~~~0.0486 \\
	\hspace*{0ex}~~~~~  5~~~~~~~~16.95~~~~~16.9986~~~~~-0.0486 \\
	\hspace*{0ex}~~~~~  6~~~~~~~~19.90~~~~~19.9957~~~~~-0.0957 \\
	\hspace*{0ex}~~~~~  7~~~~~~~~23.10~~~~~22.9928~~~~~~0.1072 \\
	\$ \textit{lfit -c x,y -v a,b -f "a*x+b" -y y line.dat --err} \\
	\hspace*{0ex}~~~~~2.99714~~~~~~2.01286 \\
	\hspace*{0ex}~~~0.0253144~~~~~0.121842 
\end{lcmd}
\caption{These pieces of commands show the two basic operations of 
\texttt{lfit}: the first invocation of \texttt{lfit} fits a straight
line, i.e. a model function with the form of $ax+b=y$ to the data
found in the file \texttt{line.dat}. This file is supposed to contain
two columns, one for the $x$ and one for the $y$ values.
The second invocation of \texttt{lfit} evaluates the model function.
Values for the model parameters ($a$, $b$) are taken from the command line
while the individual data points ($x$, $y$) are still read from the 
data file \texttt{line.dat}. The evaluation mode allows the user to 
compute (and print) arbitrary functions of the model parameters \emph{and}
the data values. In the above example, the model function itself
and the fit residuals are computed and printed, following the read
values of $x$ and $y$. Note that the printed values are formatted for
a minimal number significant figures (\%6.4g) or for a fixed number of  
decimals (\%8.2f or \%8.4f).
The last command is roughly the same as the first command for regression,
but the individual uncertainties are also estimated by normalizing
the value of the $\chi^2$ to unity.}
\label{fig:lineexample}
\end{figure*}


\begin{figure*}
\begin{lcmd}
\#!/bin/sh \\
CATALOG=input.cat~ \# name of the reference catalog \\
COLID=1~ ~ ~ ~ ~ ~ \# column index of object identifier (in the \$CATALOG file) \\
COLX=2 ~ ~ ~ ~ ~ ~ \# column index of the projected X coordinate (in the \$CATALOG file) \\
COLY=3 ~ ~ ~ ~ ~ ~ \# column index of the projected Y coordinate (in the \$CATALOG file) \\
COLMAG=4 ~ ~ ~ ~ ~ \# column index of object magnitude (in the \$CATALOG file) \\
COLCOLOR=5 ~ ~ ~ ~ \# column index of object color (in the \$CATALOG file) \\
THRESHOLD=4000 ~ ~ \# threshold for star detection \\
GAIN=4.2 ~ ~ ~ ~ ~ \# combined gain of the readout electronics and the A/D converter in electrons/ADU \\
MAGFLUX=10,10000 ~ \# magnitude/flux conversion \\
APERTURE=5:8:8 ~ ~ \# aperture radius, inner radius and thickness of the background annulus (all in pixels) \\
mag\_param=c0\_00,c0\_10,c0\_01,c0\_20,c0\_11,c0\_02,c1\_00,c1\_01,c1\_10 \\
mag\_funct="c0\_00+c0\_10*x+c0\_01*y+0.5*(c0\_20*x\^{ }2+2*c0\_11*x*y+c0\_02*y\^{ }2)+color*(c1\_00+c1\_10*x+c1\_01*y)" \\
for base in \$\{LIST[*]\} ; do \\
\ptab	fistar ~\$\{FITS\}/\$base.fits --algorithm uplink --prominence 0.0 --model elliptic $\backslash$ \\
\ptab	\ptab	--flux-threshold \$THRESHOLD --format id,x,y,s,d,k,amp,flux -o \$\{AST\}/\$base.stars \\
\ptab	grmatch --reference \$CATALOG --col-ref \$COLX,\$COLY --col-ref-ordering -\$COLMAG $\backslash$ \\
\ptab	\ptab	--input \$\{AST\}/\$base.stars --col-inp 2,3 --col-inp-ordering +8 $\backslash$ \\
\ptab	\ptab	--weight reference,column=\$COLMAG,magnitude,power=2 $\backslash$ \\
\ptab	\ptab	--triangulation maxinp=100,maxref=100,conformable,auto,unitarity=0.002 $\backslash$ \\
\ptab	\ptab	--order 2 --max-distance 1 $\backslash$ \\
\ptab	\ptab	--comment --output-transformation \$\{AST\}/\$base.trans || continue \\
\ptab	grtrans \$CATALOG --col-xy \$COLX,\$COLY --col-out \$COLX,\$COLY $\backslash$ \\
\ptab	\ptab	--input-transformation \$\{AST\}/\$base.trans --output - | $\backslash$ \\
\ptab	fiphot ~\$\{FITS\}/\$base.fits --input-list - --col-xy \$COLX,\$COLY --col-id \$COLID $\backslash$ \\
\ptab	\ptab	--gain \$GAIN --mag-flux \$MAGFLUX --aperture \$APERTURE --disjoint-annuli $\backslash$ \\
\ptab	\ptab	--sky-fit mode,iterations=4,sigma=3 --format IXY,MmBbS $\backslash$ \\
\ptab	\ptab	--comment --output \$\{PHOT\}/\$base.phot \\
\ptab	paste ~	\$\{PHOT\}/\$base.phot \$\{PHOT\}/\$REF.phot \$CATALOG | $\backslash$ \\
\ptab	lfit ~ ~--columns mag:4,err:5,mag0:12,x:10,y:11,color:\$((2*8+COLCOLOR)) $\backslash$ \\
\ptab	\ptab	--variables \$mag\_param --function "\$mag\_funct" --dependent mag0-mag --error err $\backslash$ \\
\ptab	\ptab	--output-variables \$\{PHOT\}/\$base.coeff \\
\ptab	paste ~ \$\{PHOT\}/\$base.phot \$\{PHOT\}/\$REF.phot | $\backslash$ \\
\ptab	lfit ~ ~--columns mag:4,err:5,mag0:12,x:10,y:11,color:\$((2*8+COLCOLOR)) $\backslash$ \\
\ptab	\ptab	--variables \$(cat \$\{PHOT\}/\$base.coeff) $\backslash$ \\
\ptab	\ptab	--function "mag+(\$mag\_funct)" --format \%9.5f --column-output 4 | $\backslash$ \\
\ptab	awk ~ ~ '\{ print \$1,\$2,\$3,\$4,\$5,\$6,\$7,\$8; \}' > \$\{PHOT\}/\$base.tphot \\
done \\
for base in \$\{LIST[*]\} ; do test -f \$\{PHOT\}/\$base.tphot \&\& cat \$\{PHOT\}/\$base.tphot ; done | $\backslash$ \\
grcollect - --col-base 1 --prefix \$LC/ --extension .lc
\end{lcmd}
\caption{A shell script demonstrating a complete working pipeline for
time series aperture photometry. The input FITS files are read from the 
directory \texttt{\$\{FITS\}} and their base names (without the 
\texttt{*.fits} extension) are expected to be in the 
array \texttt{\$\{LIST[*]\}}. These
base names are then used to name the files storing data obtained during
the reduction process. Files created by the subsequent calls of 
the \texttt{fistar} and \texttt{grmatch} programs are related to the
derivation of the astrometric solution and the respective files
are stored in the directory \texttt{\$\{AST\}}. The photometry centroids
are derived from the original input 
catalog (found in the file \texttt{\$CATALOG})
and the astrometric transformation (plate solution, stored in the \texttt{*.trans})
files. The results of the photometry are put into the directory \texttt{\$\{PHOT\}}.
Raw photometry is followed by the a magnitude transformation. This branch
involves additional common UNIX utilities such as \texttt{paste} and \texttt{awk}
in order to match the current and the reference photometry as well as 
to filter and resort the output after the magnitude transformation. 
The derivation of the transformation coefficients is done by the
\texttt{lfit} utility, that involves \texttt{\$mag\_funct} with the
parameters listed in \texttt{\$mag\_param}. This example features a quadratic
magnitude transformation and a linear color dependent correction 
(to cancel the effects of the differential refraction). The final light curves
are created by the \texttt{grcollect} utility what writes the individual
files into the directory \texttt{\$\{LC\}}.
More detailed examples are available on the web page of the project,
located at \texttt{http://fitsh.szofi.net}. These examples include
further possible applications and sample data as well (that have been
or are going to be published in and related to separate papers).}
\label{fig:apphotexample}
\end{figure*}


\section{Summary}
\label{sec:discussion}

In this paper we described a software package named \fitsh{} intended
to provide a complete solution for many problems related to 
astronomical image processing -- including calibration, source extraction,
astrometry, source identification, photometry, light curve processing
and regression analysis. The implementation scheme enables not only
the simple and portable processing but the easy cooperation with
other existing related data processing software packages. This
package has an open source codebase, so any details related to the
actual execution of the various tasks and algorithms can be traced.
Although the current implementation allows
the user a fast accomplishment of the previously listed exercises, there are
some features that needs to be improved or some implementational
aspects should be reconsidered. These include the cleanup of the 
code related to PSF analysis and some user interface functionality
and homogeneity (more similar syntax for some related tasks, more
sophisticated output formatting as provided by the users and so on). 
In order to be more compatible with existing software solutions,
some improvements are considered related to the mask handling and
built-in compression algorithms. 
In addition, implementation of features are also considered and
currently in testing stage that aids the processing of data that are 
not related directly to ``CCD imaging''. These include processing of
grism observations or applications for post-processing images acquired
in (near or far) infrared spectral regimes (e.g. Herschel Space Observatory).
The web page of the project (\texttt{http://fitsh.szofi.net})
also provides a forum for the users that are
open for discussions related to the \fitsh{} package.


\section*{Acknowledgements}

The current work of A.~P. and the recent development of this
software package have been supported by the ESA grant PECS~98073
and by the J\'anos Bolyai Research Scholarship of the
Hungarian Academy of Sciences. The author would like to thank
G\'asp\'ar Bakos and the HATNet project 
for the various ideas, the initial improvement
and support for earlier developments of the package. The author
also acknowledges the useful notes of the referee,
Jim Lewis. His comments help to be make the package more permeable
with other existing packages. The author thanks
also to Brigitta Sip\H{o}cz for her suggestions and bug reports. 
The recent ideas and assistance of Gy\"orgy Mez\H{o} and Kriszti\'an
Vida are also acknowledged. The author thanks to P\'eter Somogyi for
his help during the development of the project's web page. 


{}

\bsp

\label{lastpage}


\begin{thebibliography}{99}

\bibitem[\protect\citeauthoryear{Alard \& Lupton}{1998}]{alard1998}
Alard, C. \& Lupton, R. H.
1998, ApJ, 503, 325

\bibitem[\protect\citeauthoryear{Alard}{2000}]{alard2000}
Alard, C. 
2000, A\&AS, 144, 363


\bibitem[\protect\citeauthoryear{Bertin \& Arnouts}{1996}]{bertin1996}
Bertin, E. \& Arnouts, S.
1996, A\&AS, 117, 393



%
\bibitem[\protect\citeauthoryear{Chromey}{2010}]{chromey2010}
Chromey, F. R.: \emph{To Measure the Sky: An Introduction to Observational Astronomy}
2010, Cambridge University Press, The Edinburgh Building, Cambridge CB2 8RU, UK





\bibitem[\protect\citeauthoryear{Joye \& Mandel}{2003}]{joye2003}
Joye, W. A. \& Mandel, E.,
2003, Astronomical Data Analysis Software and Systems XII,
ASP Conference Series, Vol. 295, 489
(eds. H. E. Payne, R. I. Jedrzejewski, and R. N. Hook)



%

\bibitem[\protect\citeauthoryear{P\'al \& Bakos}{2006}]{pal2006}
P\'al, A., Bakos, G. \'A.,
2006, PASP, 118, 1474



\bibitem[\protect\citeauthoryear{P\'al}{2009b}]{pal2009phd}
P\'al, A.
2009b, PhD thesis (arXiv:0906.3486)

\bibitem[\protect\citeauthoryear{P\'al}{2010}]{pal2010}
P\'al, A.,
2010, MNRAS, 409, 975




\bibitem[\protect\citeauthoryear{Stetson}{1987}]{stetson1987}
Stetson, P. B.,
1987, PASP, 99, 191 

%
%
\bibitem[\protect\citeauthoryear{Stetson}{1989}]{stetson1989}
Stetson, P. B.,
1989, in Advanced School of Astrophysics, Image and Data Processing/Interstellar Dust, ed. B. Barbury, E. Janot-Pacheco, A. M. Magalh\~aes and S. M. Viegas (S\~ao Paulo, Instituto Astr\^onomico e Geof\'{\i}sico)






\end{thebibliography}
\end{document}